# TelEdge: Haptic Tele-Communication of a Smartphone by Electro-Tactile Stimulation Through the Edges

Taiki Takami[1], Izumi Mizoguchi[1], and Hiroyuki Kajimoto[1]

[1] Department of Informatics, The University of Electro-Communications, Tokyo, JAPAN

(Email: takami@kaji-lab.jp)

**Abstract ---** We present TelEdge, a novel method of remote haptic communication using electrical stimulation through the edges of the smartphone. The aim of this study is to explore communications that can be created by adding touch sensing and haptic feedback using the electrical edge display to conventional audio-visual functionality. We conducted monitoring observations and interviews during a video call between two people, presenting interactive haptic feedback.

## 1 Introduction

In the current context dominated by the widespread use of personal smartphones, remote communication has become integral to both daily life and professional endeavors. This form of communication heavily relies on auditory and visual modalities, with limited emphasis on tactile-centric media. Recognizing the significance of touch-enriched communication, various proposals have emerged [1].

Poke [2] introduced an innovative feature on smartphones that enables the transmission of emotional touches through air pressure. SansTouch [3] combined a multimodal sleeve device with a smartphone, providing a sophisticated non-contact tactile communication experience involving temperature, pressure, and light. However, challenges such as device complexity and impediments to routine smartphone usage have surfaced.

In response to these challenges, we propose the integration of electrical stimulation into smartphones, ensuring high-resolution tactile feedback without compromising regular usage. This proposal is tailored for applications in remote communication scenarios. The primary objective of this research is to investigate tactile communication facilitated by electrical stimulation on smartphones. The paper advocates a setup enabling touch sensing and tactile feedback through the edges of smartphones within the context of remote communication.

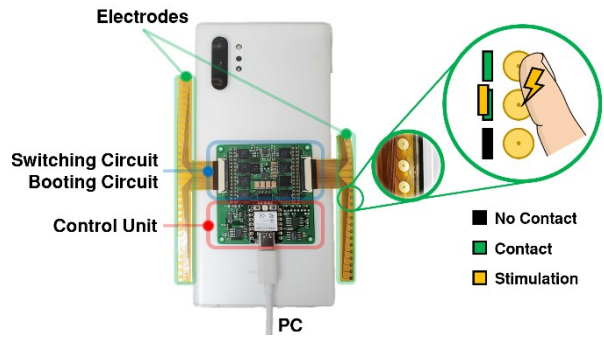

Fig.1  Electro-Tactile edge display [4]

Our previously developed electro-tactile edge display [4] delivers electrical stimuli to the fingers holding the smartphone's both edges, composed of a single row array of electrodes affixed to the edges. This approach allows for a thin, lightweight, and compact implementation, making it feasible to integrate the device into the smartphone case. Furthermore, due to its low power consumption, it eliminates the need for a special power supply. Tactile feedback delivered to the edges of the smartphone effectively reaches the fingers and palm of the holding hand. To investigate the impact of bidirectional tactile feedback during paired video calls, we conducted a monitoring experiment using the electro-tactile edge display.

## 2 Implementation

Our electro-tactile edge display [4], designed for tactile feedback and touch sensing, was used in this study. The device is equipped with 53 electrodes, with 21 on the left and 32 on the right, and is connected to a PC or smartphone via USB for power supply and communication (Fig.1 ).

Simultaneous execution of tactile feedback and touch sensing are achieved using the same electrodes, transmitting bidirectional stimulation information to the synchronized device, with the touched electrode also acting as the stimulating electrode.

### 3 MONITORING USER STUDY

We recruited four participants who had previously experienced electro-tactile stimulation, organized into two pairs for video calls (Fig.2 ). One pair consisted of friends, while the other pair comprised acquaintances who had limited conversation. This approach facilitates a comparative observation of tactile communication in close relationships versus less familiar ones.

One participant started video calls from Room A, and the other commenced the call from Room B. The interactions were systematically observed through a fixed camera to comprehend how tactile communication unfolds.

We used Discord for the video calls, each lasting 10 minutes.

As a method of bidirectional tactile feedback, electrical stimulation was applied to the overlapping positions of the participants' fingers using touch sensing (Fig.3 ). For example, if the two touches the same position, they feel the stimulation. If one holds the smartphone steadily and the other traces on the edge of the smartphone, the former feels the movement of the finger.

The electrical pulse width was 50 us, refresh rate was 60 Hz. The intensity (pulse height) of the stimuli was adjusted to a perceptible level before the experiment.

The Ethics Committee approved this study on Experiments on Human Subjects (approval number 22021) of the University of Electro-communications, Chofu, Tokyo, Japan.

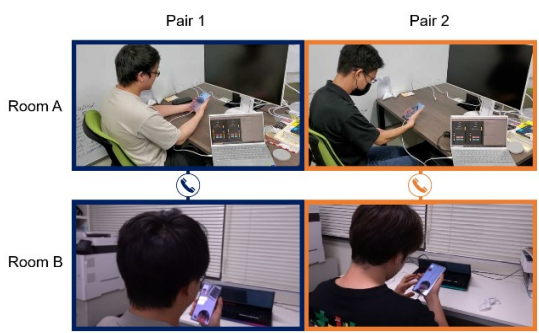

Fig.2  Scene of communication by each pair

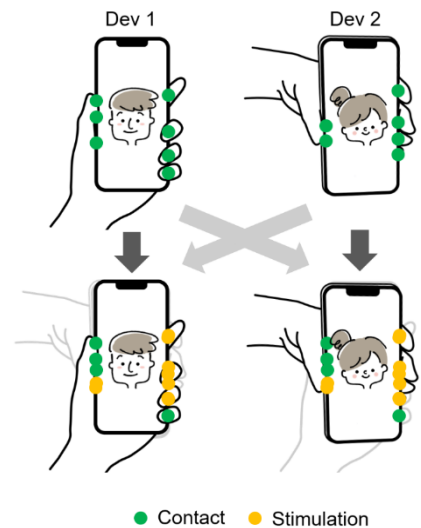

Fig.3  Electrical stimulation that is applied to the overlapping positions of touched electrodes

### 4 RESULT

In a 10-minute video call, both pairs engaged in casual conversation, repeatedly using touch gestures like tracing and focusing on tactile discussions such as "How do you feel?". During these interactions, quick finger movements were challenging to convey, while deliberate touch gestures like slow tracing or the repetitive act of gripping and releasing the smartphone akin to tapping were more effectively transmitted as sensory stimuli, receiving favorable evaluations for their intriguing nature (Fig.4 ).

Importantly, the design choice to consistently stimulate the overlapped points of contact led to uncertainty when both participants held the smartphone simultaneously. Similarly, instances were observed where simultaneous movements of fingers by both parties to convey tactile sensations resulted in ambiguous outcomes. Overall, stable communication was achieved through a method where one participant maintained a static grip while the other conveyed tactile information through finger movements.

Distinguishing differences between pairs with close relationships and those with mere acquaintances, the former used language light-heartedly, enjoying several tactile communications throughout. Conversely, the latter started with polite greetings and kind tactile communications, progressing to delightful interactions marked by shared smiles as tactile confirmations unfolded.

In post-call interviews, there was an expressed perspective that the potential lies more in entertainment than in conveying informational content through tactile communication. Acknowledging limitations in expressive capabilities with a single row edge-based tactile communication setup, suggestions were made for interactive games like the "Guess Where I'm Touching" game or simple in-game communication among friends (e.g., signaling with "GO" or "OK").

Based on the observations and interviews mentioned earlier, there is potential in incorporating tactile media as an entertainment in calls and its utility as an initial means of communication for individuals lacking prior acquaintance.

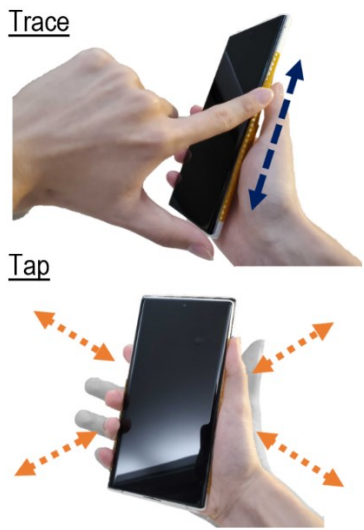

Fig.4  Touch gestures that received favorable evaluations by participants

## 5  LIMITATION AND DISCUSSION

In this paper, we used 1-D electrode arrays for electrical stimulation and it is challenging to express rich information akin to a 2-D representation accordingly. However, it's essential to note that presenting simple actions, such as "tracing" or "tapping," suggests communication potential.

## 6  CONCLUSION

In this study, electrical stimulation was employed as a tactile method at the edges of the smartphone to observe tactile communication. The results demonstrated the emergence of simple communications, such as "tracing" or "tapping, " which were enjoyable regardless of interpersonal dynamics. Considering the inherent challenges of conveying complex information, indications suggested potential applications in enhancing entertainment during remote communication, thus expanding the scope of tactile communication usage.

As a next step, we plan to conduct supplementary experiments, including applications in entertainment, particularly in scenarios with one-to-many interactions, and explore tactile communication deeply using electrical stimulation on smartphones.

## ACKNOWLEDGEMENT

This research was supported by JSPS KAKENHI Grant Number JP20K20627 and KDDI Foundation.